\documentclass[12pt,preprint]{aastex}

\newcommand \gsim{ \lower .75ex \hbox{$\sim$} \llap{\raise .27ex \hbox{$>$}} } 
\newcommand \lsim{ \lower .75ex\hbox{$\sim$} \llap{\raise .27ex \hbox{$<$}} }

\newcommand \sw{{\it Swift}}

\newcommand \ko{{\it Konus-Wind}}
\newcommand \ba{BATSE}

\begin{document}
\title{Precursors in \sw\ Gamma Ray Bursts with redshift}

\normalsize \author{D. Burlon\altaffilmark{1,5}, G. Ghrlanda\altaffilmark{1}, 
G. Ghisellini\altaffilmark{1}, D. Lazzati\altaffilmark{2}, L. Nava\altaffilmark{1,3}, 
M. Nardini\altaffilmark{4}, A. Celotti\altaffilmark{4}} 
\affil{
INAF --
Osservatorio Astronomico di Brera, via Bianchi 46, I--23807 Merate, Italy\\
Universit\`a degli Studi di Milano Bicocca, P.za della Scienza 3, I-20126 Milano, Italy}
\altaffiltext{1}{INAF --
Osservatorio Astronomico di Brera, via Bianchi 46, I--23807 Merate, Italy}
\altaffiltext{2}{JILA, University of Colorado, 440 UCB, Boulder, CO 80309-0440, USA}
\altaffiltext{3}{Univ. dell'Insubria, V. Valleggio, 11, I--22100, Como, Italy}
\altaffiltext{4}{SISSA/ISAS, Via Beirut 2/4, I-34014 Trieste, Italy}
\altaffiltext{5}{Universit\`a degli Studi di Milano Bicocca, P.za della Scienza 3, I-20126 Milano, Italy}

\begin{abstract}
We study a sample of Gamma--Ray Bursts detected by the Swift satellite
with known redshift which show a precursor in the {\it Swift}--BAT
light curve.  We analyze the spectra of the precursors and compare
them with the time integrated spectra of the prompt emission.  We find
neither a correlation between the two slopes nor a tendency for the
precursors spectra to be systematically harder or softer than the
prompt ones. The energetics of the precursors are large: on average,
they are just a factor of a few less energetic (in the source rest
frame energy range 15--150 keV) than the entire bursts.  These
properties do not depend upon the quiescent time between the end of
the precursor and the start of the main event.  These results suggest
that what has been called a ``precursor'' is not a phenomenon distinct
from the main event, but is tightly connected with it, even if,
in some case, the quiescent time intervals can be longer than 100 seconds.
\end{abstract} 


\section{Introduction}

What happens in a Gamma--Ray Burst (GRB) before the main $\gamma$--ray
prompt event is still an open question.
Koshut et al. (1995, hereafter K95) searched in the \ba\ sample for
precursors defined as pulses with a peak intensity lower than that
of the main GRB and separated from it by a quiescent phase at least as
long as the main burst duration.  They found that a precursor was
present in $\sim 3$\% of the 995 GRBs detected up to May 1994: their
duration appeared weakly correlated with those of the main GRBs and on
average shorter than that of the burst.  The spectral properties of
the precursors (based on hardness ratios) showed no relation with
those of the GRB, being both softer and harder.

Lazzati (2005; L05 hereafter) searched for precursors as weak events
{\it preceding} the \ba\ trigger.  He found them in $\sim$20\% of the
bursts.  These precursors account for only a small fraction of the
total GRB counts, and their duration show a weak correlation with that
of the burst.  In contrast to those of K95, the precursors studied by
L05 are softer than the main events.

K95 also found that the typical precursor--to--burst separation time is
$\Delta t \simeq$100 s, whereas L05 showed that most precursors were
closer to the onset of the main event (with $\Delta t \simeq$ 30 s). 
These timescales are computed in the observer frame since we do not know
the redshift for most of the \ba\ bursts.  
This difference could be due to the different definition of
precursor--to--burst separation adopted (peak--to--peak separation and
interval between the onset times for K95 and L05, respectively).  Both
analysis revealed, however, that the minimum separation is of the
order of 10 s.

The main limitations of the above studies were: 
i) the lack of known distances, that prevented to quantify 
the absolute energy of the precursor event; 
ii) the poor spectral characterization of the precursor, that
was based on low resolution spectra: typically the spectrum was
described by either the hardness ratio (K95) or by a fit to a 3--4
channel broad band spectrum (L05).

Theoretical models for precursors can be separated into three classes:
the ``fireball precursor'' (Li 2007; Lyutikov \& Blandford, 2003;
Meszaros \& Rees, 2000; Daigne \& Mochkovitch, 2002; Ruffini et al. 2001); 
the ``progenitor precursor'' models (Ramirez--Ruiz, 
MacFadyen \& Lazzati, 2002; Lazzati \& Begelman 2005) and the 
``two step engine'' model (Wang \& Meszaros 2007).  
In the first class the precursor is associated to the
initially trapped fireball radiation being released when transparency
is reached.  In the second class, based on the collapsar scenario, the
precursor is identified with the interaction of a weakly relativistic
jet with the stellar envelope. A strong terminal shock, breaking out of
the envelope, is expected to produce transient emission. In both
classes of models the precursors emission is predicted to be thermal,
characterized by a black--body spectrum.  
In the third class the collapse
of the progenitor star leads to the formation of a neutron star whose
emission would be responsible for the precursor;
subsequent accretion onto the neutron star causes its collapse onto a
black hole, originating the GRB prompt.

Since the launch of the \sw\ satellite (Gehrels et al. 2004),
redshifts have been measured for a considerable number of GRBs.
It is thus possible to investigate the rest frame properties and
calculate the energetics of any precursor associated with them.
These of course are key physical quantities for the understanding of
their origin.

A major issue however is the very definition of ``precursor'', as
there is no obvious objective criterion.  For this reason we have
considered a ``loose'' operative definition for their ``selection'',
allowing ourselves to sub--select, {\it a--posteriori}, those
events sharing some characterizing property.
Thus, in our operative approach, a precursor is defined simply 
as an initial signal for which: 
\begin{itemize}
\item the peak flux is smaller than that of the main event in
the \sw--BAT 15--350 keV light curve;  
\item the flux falls below the background level before the start of 
the main event.
\end{itemize}
Our definition is quite similar to that adopted by K95 although we do
not require the precursor to precede the bursts by a time delay at
least as long as the main burst duration and, differently from L05, we
do not impose the condition that a precursor did not trigger the
detector.
Indeed, such a definition might comprise initial pulses
with very short time separation from the start of the main event,
making them effectively undistinguishable from first pulses of the
burst prompt emission. 
But how long should the temporal separation be to let us distinguish among
the two?  Or are there, instead, other (e.g. spectral) properties
which can neatly distinguish precursors from main events,
independently of the time separation?  And are they really physically
different?  These are some of the issues we are trying to investigate.

We adopt a $\Lambda$CDM cosmology with $\Omega_\Lambda=h_0=0.7$ and
$\Omega_{\rm M}=0.3$.

\section{Already known precursors with redshift}

In the literature, there are 5 bursts with precursors and known redshift.
GRB~011121 (Piro et al. 2005) and GRB~030329 (Vanderspek et al. 2004)
show two possible precursors each, preceding the burst trigger by a
few sec.  However, these two GRBs do not match our selection criteria
as the precursor candidates occur during the rising of the main event.
GRB~050820A (Cenko et al. 2006), GRB~060124 (Romano et al. 2006) and
GRB~061121 (Page et al. 2007) have precursors which triggered \sw--BAT
and preceded the main events by several tens of sec. Although \sw--BAT
could not completely follow the main events of GRB~060124 and
GRB~050820A, due to the limited burst--mode memory buffer and to the
passage over the South Atlantic Anomaly respectively, the \ko\ data
complete the light curve allowing to study the main event.  
In Table~1 their main temporal and spectral properties and energetics are
reported.



\section{Precursors in the \sw\ sample}

We have searched for precursor activity in all GRBs with measured
redshift  detected by \sw\ up to March 2008, comprising 105 GRBs.  By
applying our definition criteria, we found 15 GRBs with a precursor, 
including the three GRBs (061121, 060124 and 050820A) already
discussed in the literature.  All of them are long GRBs, i.e. 
$T_{90} > 2$ s.  
Since GRB 070306 has two precursors, this implies a total of 16
precursors in the \sw\ sample.
%
%
We have applied the standard \sw--BAT data reduction pipeline (v.2.8)
to extract light curves and spectra for the GRBs in our sample.  
We computed the
precursor and main GRB duration $T_{90}$ from the background
subtracted 15--350 keV light curves binned at 1 s. The precursor and
main GRB spectra were obtained with the standard procedure, taking into
account the energy dependent systematic errors.
The spectral analysis was performed with {\it Xspec v11.3.2}.  The
spectra were fit with a single (PL) and a cutoff (CPL) power--law
model. For the precursors the PL model provides the best fit, i.e. the
fit with the CPL model does not statistically improve 
(at the 3$\sigma$ confidence level).

Since, in some theoretical model, the precursor emission is expected to have a thermal
origin, we also fit their spectra with a black--body (BB).  For
the 9 precursors with the largest ($>$ 12) signal--to--noise ratio S/N
(integrated over 15--150 keV), the BB representation is excluded at
more than 3$\sigma$ in 6 cases and between 2 and 3$\sigma$ in 3 cases.  In
GRB 060115 and GRB 071010B (S/N$>$10) the residuals of the BB fit show
systematic deviations at low and high energies.  For these two precursors,
an hybrid BB+PL model (Ryde 2005, but see Ghirlanda et al. 2007)
yielded a BB component contributing  $\sim$50\%
of the total flux, but this model was only 1$\sigma$ significantly
better than the single PL model.  For the remaining 5 precursors the
low S/N ($<$10) does not allow to discriminate between the BB, PL, or
other models with the same number of free parameters.

\section{Results}

\begin{figure}
\hskip -2.1 true cm
\scalebox{1.31}{\plottwo{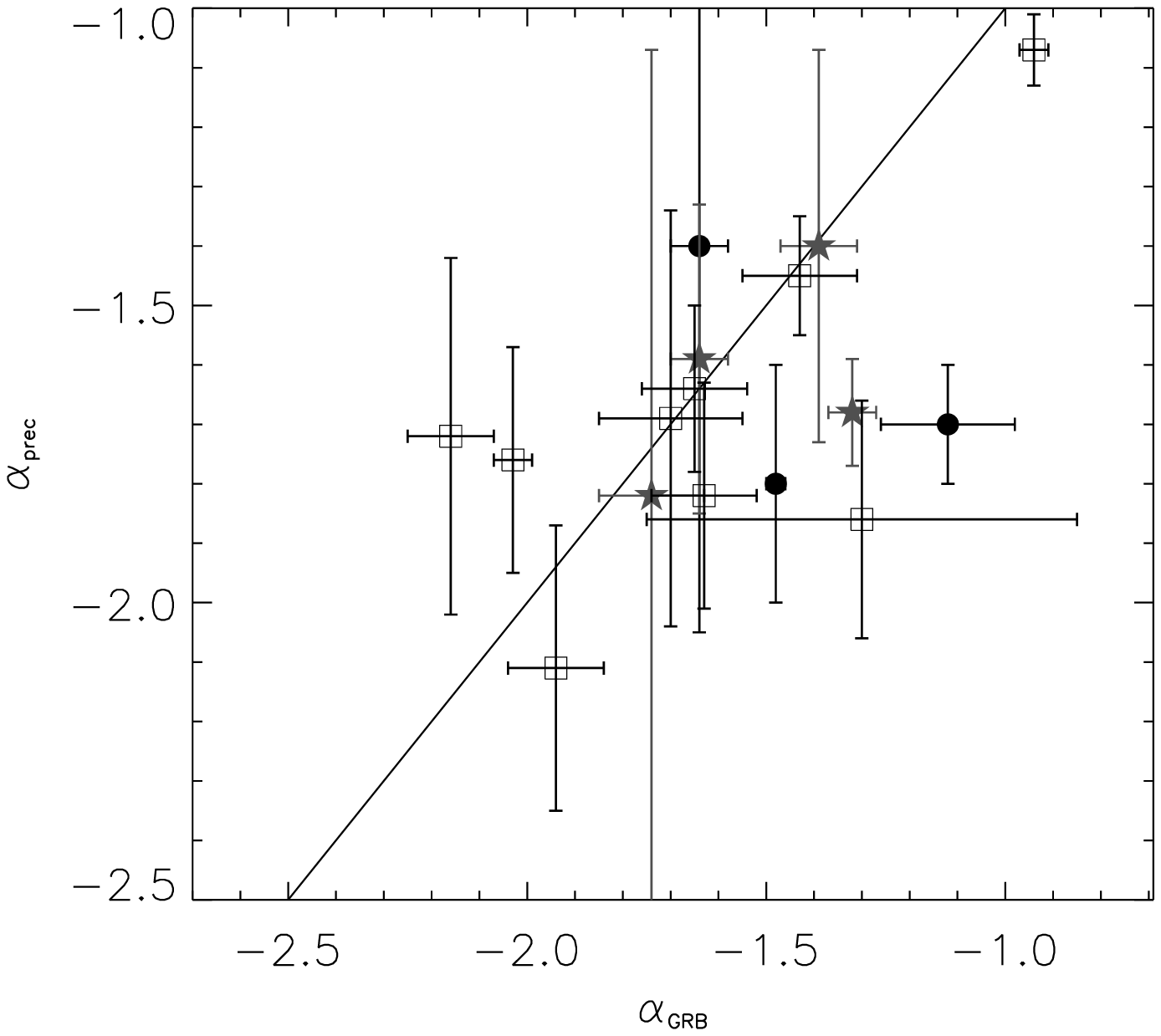}{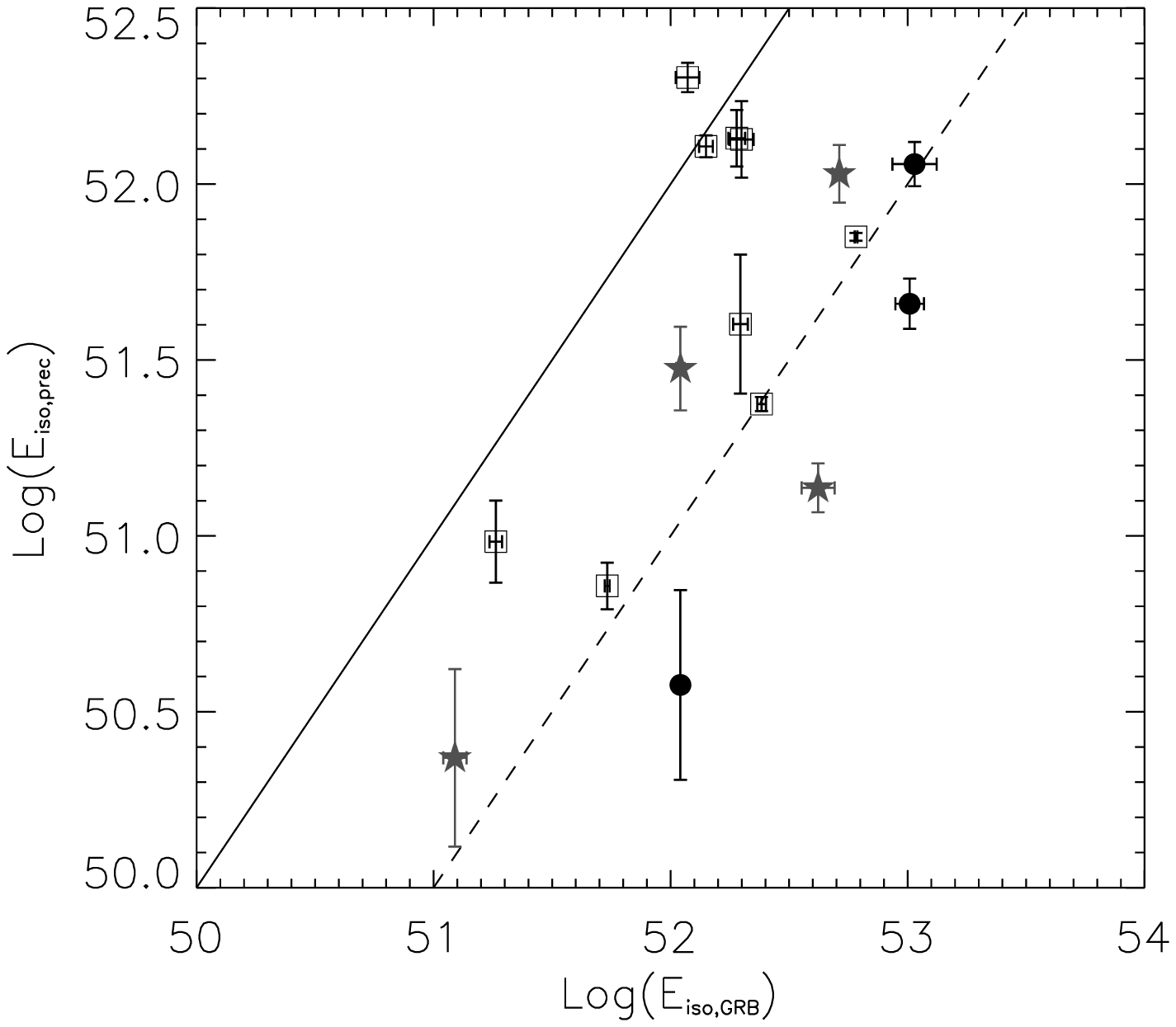}} 
\caption{ 
{\it Left panel}: 
Precursor versus burst photon spectral index. 
Different symbols correspond to
different {\it rest frame} time separation $\Delta t$, between the
precursor end and the start of the main event: 
filled circles: $\Delta t>40$ s;
grey stars: $15<\Delta t<40$ s; 
open squares: $\Delta t< 15$ s.  
{\it Right panel}: 
Precursor versus burst isotropic energy in the 15--150 keV rest frame band.  
Same symbols as in the left panel.  
The solid and dashed lines correspond to precursors having equal and 
1/10 the energetic of the main event, respectively.}
\label{alfa}
\end{figure}

In Fig.~\ref{alfa} (left panel) the photon spectral indices of
precursor ($\alpha_{\rm prec}$) and main GRB ($\alpha_{\rm GRB}$) are
compared. 
There is no clear tendency for the precursor emission to be harder or softer
than the prompt. The typical photon index
distributions of precursors and main events are both fully consistent
with that for the whole \sw\ sample recently published (Sakamoto et
al. 2008).

As all of the precursor spectra are best fitted by a single power--law,
it is not possible to determine either the peak energy in $\nu F_{\nu}$
or the bolometric energy $E_{\rm iso}$. 
As the best possible proxy for the latter, we consider 
the energy emitted, in the rest frame, between 15 and 150 keV. 
As shown in Fig.~\ref{alfa} (right panel) the precursor 
isotropic energy is on average  $\sim$1/3 of
that of the corresponding main GRB event.



In order to examine the possible role of the duration of the quiescent
time, i.e. the time delay $\Delta t$ between the end of the precursor
and the start of the main event, we have divided the sample into three
subsets, according to $\Delta t$ calculated in the source rest frame
[$\Delta t \equiv (T_{\rm 1, main} -T_{\rm 2, prec})/(1+z)$], $T_{\rm
1, main}$ and $T_{\rm 2, prec}$ are reported in Tab.~1).  $\Delta t$
is broadly distributed between a few sec and a few tens of sec with an
average value $\sim$10 s. By comparing the behavior of the
precursors belonging to the three subgroups we can check whether our
sample is ``contaminated'' by initial ``pulses'' that possibly have
properties and origin different from those of ``true'' precursors.
Both panels of Fig. \ref{alfa} -- where the events are coded according
to $\Delta t$ (i.e. $\Delta t < 15$ s, 15 s $<\Delta t<40$ s and
$\Delta t> 40 $ s) -- show that there is no clear separation among
them.
In terms of energetics, a K--S test on the distributions of $E_{\rm
iso}$ for \sw\ GRBs with redshift (adapted from Sakamoto et al. 2008)
with and without precursors indicates that they are consistent with
being drawn from the same distribution (null hypothesis probability $P$=3\%). 
As expected the corresponding distribution for the precursors 
is shifted towards lower $E_{\rm iso}$.


Finally, an analysis of the rest frame pulse durations $T_{90}$
supports the finding by L05, namely the existence of a tentative
(1$\sigma$ significant) correlation between the $T_{90}$ of precursors and
the $T_{90}$ of  main events.


\section{Conclusions}

Our results point to a clear but puzzling conclusion: the spectra and
energetics of the selected initial pulses, being them bona fide
precursors or not, are indistinguishable from those of the main event.
While this could be not surprising for ``precursors'' which were in
fact the initial pulse of the main event, in cases like GRB 060124 and
GRB 050820A the precursor precedes the main event by $\sim$100 s (rest
frame time): yet they behave as the rest of the main emission, like
``normal'' initial pulses.  

This forces us to re--consider what the very same precursor phenomenon
is.  Our finding contrasts with that by L05, who found precursors much
fainter and significantly softer than the main event.  However the
precursor selection criteria are different and in particular the
requirement by L05 that the precursors did not trigger \ba\ obviously
biased the sample against strong precursors. Our result are instead
more consistent with the Koshut et al. (1995) one, whose selection
criteria is similar to ours.
We therefore cannot exclude that there are two kinds of ``precursors'':
one as strong as and spectrally similar to the main event and the other being 
softer and dimmer.  
{\it But -- independent of that -- both can occur $\sim$100 s
before the main event.}  
Indeed, this long delay is both the most intriguing feature and the
main difficulty for all the proposed progenitor interpretation.  As
discussed by Wang \& Meszaros (2007), the progenitor class of
models cannot explain delays longer than $\sim$10 s.

The origin of quiescent times has been discussed by Ramirez--Ruiz,
Merloni \& Rees (2001), who considered the possibility that a
temporal modulation in Lorentz factor of ejected shells/relativistic
outflow would lead to time dependent emission via dissipation in
internal shocks.  ``Fireball'' models predict too short quiescent
timescales if the main prompt emission mechanism is internal shocks
taking place at typical radii $R= 10^{13}R_{13}$ cm, since $\Delta
t\sim R/(c\Gamma^2)\sim 0.03 R_{\rm 13}/\Gamma^2_2$ s.
%
%
External shocks occurring at $R\sim 10^{16}$ cm can lead to
time delays similar to what observed, but -- in the case of a
homogeneous fireball interacting with an homogeneous interstellar
medium (ISM) -- this process hardly accounts for fast prompt
variability, suppressed by the curvature effect.  More complex
external shock scenarios can overcome this problem (e.g. Dermer et
al. 1999), but in turn require a strongly clumped ISM.

A second issue emerging from our results concerns the spectral shape
of the precursor. The non--thermal appearance of the spectra is not
the chief problem, as this may arise as convolution of black-body
emission at different temperatures and/or from different locations,
consistently with the predicted thermal character. What
remains puzzling (or revealing) is that, on average, the power--law fit
spectral indices are very similar to those of the main event.
The large energetics of the precursors studied here is also difficult
to explain within the precursor models proposed so far as, whatever the
progenitor nature, they rival the main event energetics.

In the collapsar model, the precursor photons may be produced in a
region emerged from the progenitor star. Indeed, heated cocoon
material has been proposed as responsible for the precursor
(Ramirez-Ruiz, McFadyen \& Lazzati 2002), but the expected energetics
would be low compared to our findings.  This also applies to the
scenario proposed by Lazzati \& Begelman (2005), where the jet opening
angle increases in time, so an observer off axis could detect the
prompt emission after the precursor, when the jet angle becomes equal
to the viewing angle.
The ``two steps'' engine model (Wang \& Meszaros 2007) envisages that
the precursor is associated with the cooling phase of the
proto--neutron star and the delay time should correspond to the
accretion phase which ultimately leads to the collapse of the neutron
star to a black hole, when the ``normal'' GRB activity begins. 

An alternative possibility is that precursors do not represent any
distinct physical process, but are simply a manifestation of the same
phenomenon producing the prompt emission, which sometimes does give
raise to quiescent intervals between emission peaks.  
We can put an upper limit on the energy emitted during these quiescent times
considering that the BAT sensitivity for a 5$\sigma$ detection is
$2\times 10^{-10} (dT/20~ {\rm ks})^{-0.5}$ erg cm$^{-2}$ s$^{-1}$
(Markwardt et al. 2007), where $dT$ is the exposure time.  Using the
delay times reported in Tab. 1 we estimate the mean value of the
1$\sigma$ upper limit to the energy emitted during the quiescence. The
ratio of these limits to the precursors energy ranges from 0.012 for
GRB~061007 to 0.25 for GRB~060124, with an average of 0.14.

Finally, we applied the same selection criterion, adopted for
 precursors, to search for emission episodes (``postcursors'')
 following the main bursts and separated by a quiescent phase. Within
 the 15 GRBs with precursors, GRB~060210 and GRB~0508020A show two and
 three pulses after the main burst\footnote{We recall 
that for GRB~050820A, \sw\ entered the {\it SAA}
  during the main event and therefore we could not perform the
  spectral analysis of the BAT data.}.  For GRB~060210 the two
postcursors (separated by 60 and 150 s from the end of the main burst)
have spectral indices $-1.76 \pm 0.28$ and $-1.83\pm0.39$ and
energetics $(7.31 \pm 2.29)\times 10^{51}$ erg and
$(5.04\pm2.14)\times 10^{51}$ erg, respectively. The spectra are 
softer and the energetics smaller than the main event and
the precursor. Since this is the only burst, in our sample, having
both a precursor and a postcursor, we cannot draw any strong
conclusion. We plan to study spectra and energetics of postcursors by 
relaxing the condition of having also a precursor in a forthcoming
paper.

\acknowledgements 
We thank the referee for constructive comments. 
This research was partly supported by PRIN--INAF 2008 and ASI
I/088/06/0 grants. We acknowledge the use of public data from the 
\sw\ data archive.

\clearpage

\setcounter{table}{0}
\begin{table*} 
\hskip -1.7 true cm
\begin{footnotesize}
\begin{tabular}{|llllllllllll|}

\hline 
GRB       &~~$z$   & \vline   & &&  Main pulse  & &\vline & & & Precursors  & \\
          &      & \vline    &$T_1$ &$T_2$     &~~~~~~$\alpha$ &~~~~~~$E_{\rm iso}^a$ &\vline&$T_1$ &$T_2$ &~~~~~~$\alpha$ &~~~~~~$E_{\rm iso}^a$ \\ 
          &      &\vline    &s     &s         &         &~~~~~~erg             &\vline&s   &s       &        &~~~~~~erg  \\
\hline
050820A$^*$ &2.6   & \vline   &225     &553  &--1.12$^{+0.13}_{-0.15}$      &1.07($\pm$0.23)E53$^a$&\vline
                                &--17    &22       &--1.74$\pm$0.08               &1.14($\pm$0.16)E52$^a$\\
060124$^+$ &2.297 & \vline    &301.2   &811.2&--1.48$\pm$0.02               &1.02($\pm$0.14)E53$^b$&\vline
                                &--1.5   &13.5     &--1.80$\pm$0.20               &4.57($\pm$0.75)E51$^b$\\
061121$^{+ \#}$   &1.314 & \vline    &61.8    &83.38&--1.32$\pm$0.05               &4.19($\pm$0.67)E52$^c$&\vline
                                &--5     &10       &--1.68$\pm$0.09               &1.37($\pm$0.22)E51$^d$\\ 
\hline
071010B &0.947   &\vline    &--1.5   &15.23    &--2.03$\pm$0.04               &5.40($\pm$0.13)E51 &\vline
                            &--30.0  &--12.77  &--1.76$\pm$0.19               &7.21($\pm$0.11)E51\\
070411  &2.954   &\vline    &49.3    &98.3     &--1.65$\pm$0.11               &1.41($\pm$0.93)E51 &\vline
                            &--19.7  &31.3     &--1.64$\pm$0.14               &1.28($\pm$0.09)E52\\
070306  &1.49    &\vline    &83.5    & 154.5   &--1.64$\pm$0.06               &1.10($\pm$0.05)E52 &\vline
                            &--118.5 &--103.5  &--1.40$\pm$0.65               &3.77($\pm$2.34)E50\\
        &        &\vline    &        &         &                              &                   &\vline
                            &--12.48 &40.51    &--1.59$\pm$0.26               &2.99($\pm$0.82)E51\\
061007  &1.261   &\vline    &27.2    &71.2     &--0.94$\pm$0.03               &6.06($\pm$0.08)E52 &\vline
                            &--2.8   &12.2     &--1.07$\pm$0.06               &7.08($\pm$0.18)E51\\
060729  &0.54    &\vline    &56.9    &123.9    &--1.74$\pm$0.11               &1.23($\pm$0.14)E51 &\vline
                            &--1.1   &29.4     &--1.8$\pm$20.75               &2.34($\pm$1.36)E50\\
060714  &2.711   &\vline    &69.9    &116.9    &--1.30$\pm$0.47               &1.90($\pm$0.16)E52 &\vline
                            &--13.1  &43.9     &--1.86$\pm$0.20               &1.35($\pm$0.25)E52\\
060707  &3.425   &\vline    &--7.3   &49.7     &--1.70$\pm$0.15               &1.97($\pm$0.14)E52 &\vline
                            &--48.3  &--23.3   &--1.69$\pm$0.35               &4.00($\pm$1.82)E51\\
060210  &3.91    &\vline    &--72.3  &21       &--1.39$\pm$0.08               &5.15($\pm$0.34)E52 &\vline
                            &--236.3 &--200.3  &--1.40$\pm$0.33               &1.07($\pm$0.20)E52\\
060115  &3.53    &\vline    &78.9    &129.9    &--1.63$\pm$0.11        	      &1.99($\pm$0.24)E52 &\vline
                            &--22.1  &30.9     &--1.82$\pm$0.19               &1.34($\pm$0.34)E52\\
050401  &2.90    &\vline    &20.7    & 29.71   &--1.43$\pm$0.12               &1.18($\pm$0.14)E52 &\vline
                            &--7.28  &6.71     &--1.45$\pm$0.10        	      &2.01($\pm$0.20)E52\\
050318 &1.44     &\vline    &22.9    &29.9     &--1.94$\pm$0.09               &1.83($\pm$0.11)E51 &\vline
                            &--1.1   &5.8      &--2.11$\pm$0.24               &9.63($\pm$2.59)E50\\
050315  &1.949   &\vline    &--6.4   &52.      &--2.16$\pm$0.09               &2.42($\pm$0.10)E52 &\vline
                            &--57.5  &--25.5   &--1.72$\pm$0.30               &2.37($\pm$0.11)E51\\
\hline
\end{tabular}
\caption{Data from Swift/BAT except for: $^*$ Konus-Wind; $^+$
Konus--Wind, precursor from Swift.  
$T_1$ and $T_2$ are in the observer frame.  
$E_{\rm iso}$ is computed in the 15--150
keV rest frame band, except for $^*$(20-1000 keV), $^+$(20-2000 keV)
and $^\#$(20-5000 keV).  
Peak energies $E_{\rm peak}$ (keV) of main
pulses: 367$^{+95}_{-62}$ (050820A); 193$^{+78}_{-39}$ (060124);
557$\pm$66 (061121); 41.0$\pm$8.5 (060714). This corresponds to a
cut--off power-law model for all bursts. Errors are given at $90\%$
confidence level.  References: (a) Cenko et al. (2006); (b) Romano et
al. (2006); (c) Ghirlanda et al. (2008, and references therein); (d)
Page et al. (2007).  }
\label{tab1}
 \end{footnotesize}
\end{table*}

\clearpage

\begin{thebibliography}{}
\bibitem[]{} Cenko S.B., et al., 2006, ApJ, 652, 490
\bibitem[]{} Daigne F., Mochkovitch R., 2002, MNRAS, 336, 1271
\bibitem[]{} Dermer C., B\"ottcher M. \& Chiang J., 1999, ApJ, 515, L49 
\bibitem[]{} Gehrels N., et al., 2004, ApJ, 611, 1005
\bibitem[]{} Ghirlanda G., Bosnjak Z., Ghisellini G., Tavecchio F. \& Firmani C., 
             2007, MNRAS, 379, 73
\bibitem[]{} Ghirlanda G., Nava L., Ghisellini, G., Firmani C. \& Cabrera J.I.,
             2008, MNRAS, in press (arXiv:0804.1675)
\bibitem[]{} Koshut T., et al., 1995, ApJ, 452, 145 (K95)
\bibitem[]{} Lazzati D., 2005, MNRAS, 357, 722 (L05)
\bibitem[]{} Lazzati D. \& Begelman M.C., 2005, ApJ, 629, 903
\bibitem[]{} Li L.--X., 2007, MNRAS, 380, 621
\bibitem[]{} Lyutikov M. \& Blandford R.D., 2003, preprint (astro-ph/0312347)
\bibitem[]{} Markwardt C.B., 
  et al., 2007, ``The Swift BAT Software Guide'' (v. 6.3) (swift.gsfc.nasa.gov/docs/swift/analysis)
\bibitem[]{} Meszaros P. \& Rees M.J., 2000, ApJ, 530, 292
\bibitem[]{} Page K.L., et al., 2007, ApJ, 663, 1125
\bibitem[]{} Piro L., et al., 2005, ApJ, 623, 314
\bibitem[]{} Ramirez--Ruiz E., Merloni A. \& Rees M.J., 2001, MNRAS, 324, 1147
\bibitem[]{} Ramirez--Ruiz E., MacFadyen A.I. \& Lazzati, 2002, MNRAS, 331, 197
\bibitem[]{} Romano P. et al., 2006, A\&A, 456, 917
\bibitem[]{} Ruffini R., Bianco C.L., Fraschetti F., Xue S.-S., Chardonnet P., 2001, 
             ApJ, 555, L113
\bibitem[]{} Ryde F., 2005, ApJ, 625, 95   
\bibitem[]{} Sakamoto T., et al., 2008, ApJS, 175, 179
\bibitem[]{} Vanderspek R., et al. 2004, ApJ, 617, 1251
\bibitem[]{} Wang X.--Y. \& Meszaros P., 2007, ApJ, 670, 1247

\end{thebibliography}
\end{document}